\documentclass[a4paper,12pt]{article}
\usepackage{amsmath}
\usepackage{amsfonts}
\usepackage{amssymb}
\usepackage[dvips]{graphicx}

\title{Decomposition of variance in terms of conditional means}

\author{
Alessandro Fig\`a Talamanca \thanks{Dept. of Mathematics, University of Rome ''La Sapienza'', sandroft@mat.uniroma1.it}
\and Angelo Guerriero \thanks{Alma Laurea, guerrier@stat.unibo.it}
\and Alberto Leone \thanks{Alma Laurea, alberto.leone@almalaurea.it}
\and Gian Piero Mignoli \thanks{Alma Laurea, gianpiero.mignoli@almalaurea.it}
\and
Enrico Rogora \thanks{Dept. of Mathematics, University of Rome ''La Sapienza'', rogora@mat.uniroma1.it}
}

\date{}

\begin{document}

\parindent=0cm

\maketitle

\begin{abstract}
We test against two different sets of data an apparently new
approach to the analysis of the variance of a numerical variable
which depends on qualitative characters. We suggest that this
approach be used to complement other existing techniques to study the
interdependence of the variables involved. According to our method
the variance is expressed as a sum of orthogonal components,
obtained as differences of conditional means, with respect to the
qualitative characters. The resulting expression for the variance
depends on the ordering in which the characters are considered. We
suggest an algorithm which leads to an ordering which is deemed
natural. The first set of data concerns the score achieved by a
population of students, on an entrance examination, based on a
multiple choice test with 30 questions. In this case the qualitative
characters are dyadic and correspond to correct or incorrect answer
to each question. The second set of data concerns the delay in
obtaining the degree for a population of graduates of Italian
universities. The variance in this case is analyzed with respect to
a set of seven specific qualitative characters of the population
studied (gender, previous education, working condition, parent's
educational level, field of study, etc.)
\end{abstract}

\section{Introduction and methodology}\label{intro}
Let $X= (x_1, \dots x_N)$ be a numerical variable defined on a
population $P$ of $N$ individuals. We may think of $X$ as an element
of a real vector space $   \mathbf{L}$ of dimension $N$. We equip $
\mathbf{L}$ with a real, normalized scalar product: for $X, Y \in
\mathbf{L}$, and $Y=(y_1, \dots y_N)$, we define:
$$
<X,Y> = \frac{1}{N}\sum_{i=1}^N x_i y_i.
$$
The length or norm of a vector is defined in terms of the scalar
product:
 $$
 \|X\|^2= <X,X>.
 $$
The mean value of a vector $X$ is of course the scalar
$$
\overline{X}= \frac{1}{N}\sum_{i=1}^N x_i.
$$
We may also think of the mean value as a vector $E_0(X)$ of $
\mathbf{L}$ having all its components equal to the scalar
$\overline{X}$. In this context $E_0$ may be thought of as a linear
operator defined on $ \mathbf{L}$ and mapping $ \mathbf{L}$ into the
subspace of constant vectors. The variance of $X$ can be written
then as:
$$
V(X) = \|X-E_0(X)\|^2 = < X-E_0(X), X-E_0(X)>.
$$
We now suppose that the indices $i=1, \dots, N$, correspond to
individuals of a population $P$, and that $X$ is a numerical
variable defined on the population $P$. We further suppose that
$\pi$ is a partition of the population $P$ into $q$ disjoint classes
$P_1, P_2, \dots, P_q$. Denote by $|P_j|$ the number of elements of
$P_j$, so that $N = |P_1|+ \dots +|P_q|$. We can then define a
vector $E_\pi(X)$ with components:
\begin{equation}
E_\pi (X)_i= \frac{1}{|P_k|} \sum_{j \in P_k} x_j \qquad  (i \in
P_k).
\end{equation}
Observe that two components of this vector are identical if their
indices belong to the same class $P_k$ of the partition $\pi$. The
trivial identity:
$$
X-E_0(X)= E_\pi(X)- E_0(X) + X-E_\pi(X),
$$
implies
$$
 V(X)= \|X-E_0(X)\|^2= \|E_\pi(X)-E_0(X)\|^2 + \|X-E_\pi(X)\|^2,
$$
because, as it is easily seen, $E_\pi(X)-E_0(X)$ and $X-E_\pi(X)$
are orthogonal vectors.

Suppose now that $\pi_1, \pi_2, \dots, \pi_n$ is a finite sequence
of partitions of the population $P$, into respectively $q_1, q_2,
\dots, q_n$, classes. Suppose further that each partition $\pi_j$ is
a refinement of the partition $\pi_{j-1}$. (This means that each
class of the partition $\pi_{j}$ is contained in a class of the
partition $\pi_{j-1}$). Define for completeness the trivial
partition $\pi_0$ consisting of the full population $P$. Let
$P_k^j$, for $k=1, \dots q_j$ be the disjoint classes of the
population $P$ relative to the partition $\pi_j$. With reference to
the partition $\pi_j$ define the operator
$$
E_j(X)= E_{\pi_j}(X).
$$
In this fashion (1) reads:
 $$
E_{\pi_j}(X)_i = \frac{1}{|P_k^j|} \sum_{h\in P_k^j} x_h, \quad (i
\in P_k^j).
$$

 Observe that this definition makes
sense also in the case $j=0$. The trivial identity
\begin{equation}\label{dec}
X-E_0(X)= \sum_{j=1}^n [E_j(X)-E_{j-1}(X)] + X-E_n(X),
\end{equation}
implies, because of the orthogonality of the terms on the right hand
side of (\ref{dec}),
\begin{equation}\label{V.dec}
V(X) = \sum_{j=1}^n \|E_j(X)-E_{j-1}(X)\|^2 + \|X-E_n(X)\|^2.
\end{equation}
We are interested in the case in which the sequence of partitions
$\pi_j$ is defined by a sequence of qualitative characters $C_1,
C_2, \dots, C_n$ of the population $P$. We can define the partition
$\pi_j$ by considering the classes of the population formed by
individuals with identical values of the first $j$ characters.

In this case the first $n$ summands on the right hand side of
(\ref{V.dec}) represent the contributions to the variance of the $n$
qualitative characters $C_1, \dots, C_n$ within the population
considered.

Observe however that, while the sum of the first $n$ terms of the
right hand side of (\ref{V.dec}) is independent of the order in
which the characters $C_1, \dots, C_n$ are considered, the operators
$E_j$, for $0 < j < n$ are defined with respect to partitions which
strongly depend on the order in which the characters are taken. As
an obvious consequence, the value of each term $
\|E_j(X)-E_{j-1}(X)\|^2$ also depends on the order of the
characters. In a different order the characters would define a
different set of partitions; only $\pi_0$ and $\pi_n$, and
consequently $E_0$ and $E_n$ are independent of the chosen order.

We are led therefore to look for a natural order of the qualitative
characters considered. We propose an ordering based on systematic,
step by step, comparisons of the conditional means with respect to
the variables considered. This ordering, which we call
\textsl{Stepwise Optimal Ordering} (SOO) is defined as follows:

We choose the character $C_{1}$ and the corresponding partition
$\pi_1$ which maximizes $ \|E_1(X)-E_0(X)\|^2$. If $C_{1}, \dots,
C_{k}$ are chosen, the character $C_{k+1}$ is chosen so that it
refines the partition $\pi_k$ into the partition $\pi_{k+1}$ in such
a way that the value $ \|E_{k+1}(X)-E_{k}(X)\|^2$ is largest.

The order $C_1, \dots, C_n$ determined in this fashion may be
considered as a ranking of the variables. One should be aware,
however, that this ranking cannot be interpreted in terms of
relative importance in determining the phenomenon measured by the
variable $X$. As will be seen in the applications below, the
qualitative characters considered may be far from independent. This
may imply that a character which is recognized as a primary cause of
the intensity of the phenomenon measured by $X$, may be mediated by
other characters to whom it is associated, and therefore appear in
the last positions of the ranking.

We do not propose a clear cut interpretation of the significance of
the ranking obtained by our method, nor of the relative size of the
first $n$ addends which appear in (\ref{V.dec}), when the
qualitative characters are ordered according to our prescription. On
the contrary, rather than expecting straight answers, we expect that
both the ranking and the relative size of the addends in the
expression (\ref{V.dec}) would solicit questions concerning the
dependence of the variable $X$ on the qualitative variables and the
interdependence of the qualitative variables themselves (with all the 
cautions regarding the possibility to consider causal relations 
between the variables, \cite{blalock,sobel96,sobel98,sobel00}).

Nevertheless, in the very special case considered in the simulated
experiment of \mbox{Section \ref{Sec.4}}, our method yields a
ranking that reflects the relative weight of the characters.

In the following two sections we apply our method and discuss the
"ranking" of the qualitative characters, thus obtained to the two
sets of data mentioned in the abstract. The fourth section is
dedicated to a simulated experiment.

We should mention that the ideas contained in Chapter 8D of \cite{D}
were influential in the inception of this work, which started as an
attempt to apply Diaconis' ideas to the case of tree-structured
data, under the action of the group of tree-automorphisms. Under
this action the ranges of the operators $E_j-E_{j-1}$ turn out to be
irreducible subspaces of $ \mathbf{L}$.
\section{The score on an entrance examination}

Entering students of the University of Rome "La Sapienza" in
scientific and technical fields take a multiple choice test in
mathematics, which consists of $30$ questions. The test, in Italian,
may be downloaded at \cite{test}. At the moment the purpose of the
test is to discourage students who do not have an adequate
background, and to make students aware of their potential
weaknesses.

We consider a population of $2,451$ students who took the test in
2005, and we let $X$ be the score achieved by each student, that is
the number of correct answers. The variable $X$ depends on the $30$
dyadic characters, corresponding to the correct or incorrect answer
to each question. Of course,  in this case, $E_{30}(X)=X$, and
$$
V(X) = \|X-E_0(X)\|^2 = \sum_{j=1}^{30} \|E_j(X)- E_{j-1}(X)\|^2.
$$

The variable $X$ takes values between $0$ and $30$. Its mean value
is $12.9$ and the variance is $V(X) = 29.8$. The histogram of $X$ is
in Fig. 1.

\begin{center}
\begin{figure}
\includegraphics[width=8cm,angle=270]{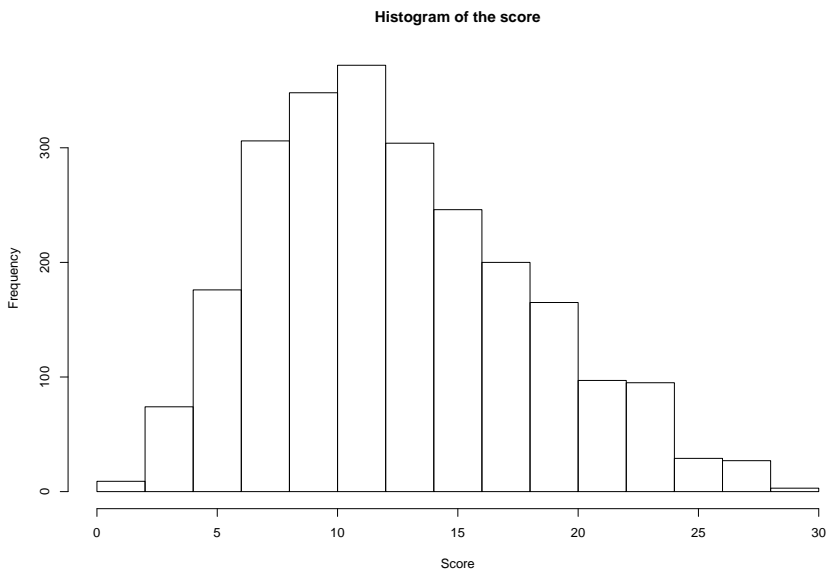}
\caption{\label{score} The histogram of the
score.}
\end{figure}
\end{center}

An application of our method shows that just ten questions, chosen
according to the ranking we propose, "explain" $88\%$ of the
variance. In other words, if we write
$$
V(X) = \sum_{j=1}^{10} \|E_j(X)-E_{j-1}(X)\|^2 + \|X-E_{10}\|^2,
$$
the remainder term $\|X-E_{10}\|^2= 3.58$ amounts to just $12\%$ of
$V(X)= 29.8$.

We presently list the remainders $\|X-E_k(X)\|^2$, for $k=1, \dots,
10$, obtained by applying our method, as percentage of $V(X)$.
To wit the values $c_k=\|X-E_k(X)\|^2/V(X)$,
$$
\begin{array}{ccccc}
c_1=\frac{75}{100},&c_2=\frac{59}{100},&c_3=\frac{48}{100},&
c_4=\frac{40}{100},&c_5=\frac{34}{100},\\
&&&&\\
c_6=\frac{29}{100},&
c_7=\frac{25}{100},&c_8=\frac{20}{100},&c_9=\frac{16}{100},&c_{10}=\frac{12}{100}.
\end{array}
$$

We do not claim, of course, that our method necessarily chooses the
$10$ characters for which $\|X-E_{10}(X) \|^2$ is lowest. In
general, with arbitrary data, this may not be the case.

However, in this particular case, our choice compares well with
other possible choices, as shown by the experiment which we
presently describe. We selected, at random, $300$ subsets of ten
elements of the original thirty questions and we computed the
conditional mean $E_{\pi}(X)$ with respect to the partition $\pi$
obtained by grouping together the students with identical
performance on each of the ten question chosen. We computed then
\begin{equation}
 \|X-E_{\pi}(X)\|^2,
\end{equation}
relative to each ten element choice. The results are summarized in
Fig. 2.

\begin{center}
\begin{figure}
\includegraphics[width=8cm,angle=270]{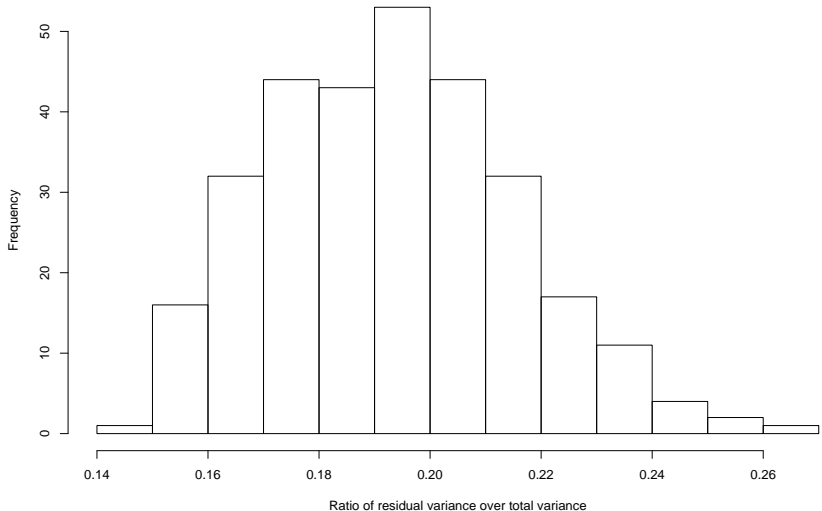}
\caption{\label{simul} Histogram of the values of residual variance
(4), as percentage of total variance for $300$ randomly selected
subsets of $10$ questions.}
\end{figure}
\end{center}

Observe that the lowest value of quantity (4) achieved by one of the
$300$ subsets we selected, is higher than $0.14$, while with our
choice of a subset of characters, we achieved a value of $0.12$.

The experiment shows that the algorithm we propose performs
decidedly better than a random choice if we want to choose ten out
of thirty questions, in such a way that the total variance of the
variable $X$ is best explained. In conclusion there is at least some
experimental evidence that our method may be used to select a small
number of characters which account for most of the variance.

It is interesting to compare our results with the results obtained with linear regression.
We found that the agreement between the results is strong.
Eight of the ten variables selected by SOO are among the ten most important
variables in terms of linear regression. Furthermore, the order of the first
five variables coincides with both methods.
We also observed that the variables selected according to SOO have the properties of
discriminating the students (the differences of percentages
of correct and incorrect answers is small).

\section{The variable "delay in completing a degree"}
The Italian  system of higher education is characterized by the
marked difference between the time employed by most students to
complete a degree and the number of years formally required to
graduate. The average delay in completing a degree is well above two
years for most fields of study \footnote{The recent reform of the
university system may hopefully change this in the near future.}. In
this section we consider a population of Italian university
graduates obtained using the data bank ``AlmaLaurea'' which
collects data of university graduates from a set of Italian
universities\footnote{AlmaLaurea Consortium is an association of 49 Italian
universities which, since 1994 collects statistical
data about the scholastic and employment records of university graduates\cite{cammelli1,cammelli2}.
The data bank of AlmaLaurea is also made available,
under certain conditions,  to prospective employers.}. The population amounts to 58,091 graduates of 27
universities in 2003. On this population the variable $X$ represents
the delay in completing the degree, computed in years, starting from
a conventional date (November 1st) in which according to formal
regulations the degree should have been completed. We excluded
delays above ten years, which should be better interpreted as
leaving and resuming the studies after several years. We study the
dependence of $X$ on seven possible characters, which are the
following:
\begin{description}
\item{(UN)} University where the degree was obtained

\item{(PE)} Parent's level of education

\item{(HS)} Type of high school attended

\item{(GD)} Grade in the final year of high school

\item{(MA)} Degree major

\item{(WO)} Working or not working during the studies

\item{(GN)} Gender
\end{description}

Proceeding as outlined in the introduction, we obtain the following
ranking of the seven variables:
\begin{center}
GD, UN, MA, HS, PE, WO, GN.
\end{center}
Accordingly we consider the operators
$$E_0, E_{1}, E_{2}, E_{3}, E_{4}, E_{5}, E_{6}, E_{7},$$ and write
\begin{equation}
V(X)= \sum_{j=1}^7 \|E_j(X)-E_{j-1}(X)\|^2 + \|X - E_7(X)|^2
\end{equation}

The variance of the variable $X$ is $V(X) = 4.61$, while the
residual variance, not "explained" by the qualitative variables
under consideration is $\|X-E_7(X)\|^2= 1.94$. The decomposition of
the variance (\ref{V.dec}) is:
$$
4.61 = (0.30 + 0.28 + 0.49 + 0.45 + 0.51 + 0.33 + 0.31)+ 1.94 = 2.67
+ 1.94
$$
Thus $2.67$ represents the portion of the variance which is
"explained" by the characters considered. We may say, therefore,
that these characters explain $62\%$ of the variance.

In this case the ranking obtained by our method is relatively
"robust". Indeed if we omit consideration of one of the characters,
the relative ranking of the other characters remains unchanged. We
do not claim of course that this type of "robustness" is inherent in
our method. It may very well occur, with different data, that
omitting one character would determine a change in the order of the
remaining characters.

We compared our results with the results obtained by using the binomial logistic regression. The delay in obtaining the degree
becomes dicotomic assigning value zero to the population of graduates with a delay less than one year ($34.1\%$) and value one
to the others ($65.9\%$). The results of our computations are shown in Table 1.

\begin{table}
\begin{center}
\begin{tabular}{l|c|r}
\textbf{Variable}&\textbf{Variance}&\textbf{var, GD=100}\\
\hline
GD&0.0119&100.0\\
UN&0.0076&63.6\\
MA&0.0097&81.7\\
HS&0.0036&30.2\\
PE&0.0012&10.3\\
WO&0.0010&8.4\\
GN&0.0001&0.5
\end{tabular}
\end{center}
\caption{Binomial logistic regression of the seven variables. In the column ``Variance'' is computed the variance of probability variation.}
\end{table}

We observe that also in this case the rank in terms of size of the variances coincides, except for one inversion, with the ranking obtained by
SOO. It should be noted however that the application of binomial logistic regression implies
an arbitrary dicotomization of the variable. Moreover, it is questionable in
this case that the binomial logistic regression would add information of
inferential value, because its application leads to many classes which are empty
or with few individuals.

\begin{center}
\begin{figure}
\includegraphics[width=8cm,angle=270]{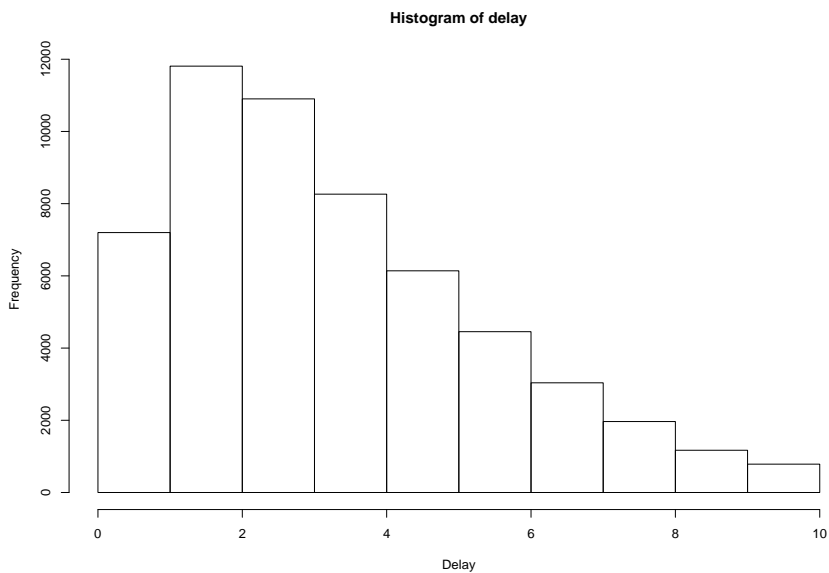}
\caption{\label{ritardo} Histogram of the
delay.}
\end{figure}
\end{center}

\section{A simulated experiment}\label{Sec.4}
In order to better understand the properties of our Stepwise Optimal
Order, we performed a simulation, repeating $20$ times the following
experiment.

First we constructed $10$ vectors $ x_1, \dots, x_{10}$ each of
$100$ components and each component extracted from a simulated
Bernoulli variable. Then we considered the variable
\begin{equation}\label{*}
x=c_1x_1+c_2x_2+\dots+c_{10}x_{10}+ \epsilon
\end{equation}
with $c_1=1$, $c_2=0.9,\dots, c_{10}=0.1$ and $\epsilon$ consisting
of $100$ independent realizations of a simulated Gaussian variable
with mean $0$ and standard deviation $0.03$.

In $18$ cases out of the $20$ observed experiments, SOO was exactly $1,2,3,\dots,10$, i.e. for the variable $x$ this order
reflected, most of the time, the size of the coefficients $c_1,
\dots, c_{10}$ which enter formula (\ref{*}). In the remaining two case the difference between SOO and the increasing order
was just one inversion.

\end{document}